# METASTABLE BOUND STATES OF THE INTERACTING TWO-DIMENSIONAL MAGNETOEXCITONS

S.A. Moskalenko[1], P.I. Khadzhi[1], I.V. Podlesny[1], E.V. Dumanov[1], I.A. Zubac[1] and M.A. Liberman[2]

[1]*Institute of Applied Physics of the Academy of Sciences of Moldova, Academic str. 5, Chisinau, MD-2028 Republic of Moldova*
[2]*Nordic Institute for Theoretical Physics (NORDITA) KTH and Stockholm University, Roslagstullsbacken 23, Stockholm, SE-106 91 Sweden*

## Abstract

The possible formation of two-dimensional (2D) magnetic bi-excitons composed of two 2D magnetoexcitons with electrons and holes on the lowest Landau levels (LLLs), with opposite center-of-mass wave vectors $\vec{k}$ and $-\vec{k}$ and with antiparallel electric dipole moments perpendicular to the corresponding wave vectors was investigated. Two spinor structures of two electrons and of two holes were considered. In the singlet-singlet state the spins of two electrons as well as the effective spins of two holes create the combinations with the total spin $S = 0$ and its projection on the magnetic field $S_z = 0$. The triplet-triplet state corresponds to $S = 1$ and $S_z = 0$. Two orbital Gaussian variational wave functions depending on $|\vec{k}|$ and describing the relative motion of two magnetoexcitons inside the molecule were used. It is shown that in the LLLs approximation the stable bound states of bi-magnetoexcitons do not exist. A metastable bound state for the triplet-triplet spin configuration a metastable bound state with the orbital wave function, having the maximum on the in-plane ring was revealed. The metastable bound state has an energy activation barrier comparable with the magnetoexciton ionization potential and gives rise to the new luminescence band due to the metastable bi-exciton-para exciton conversion with the frequencies higher than those of the para magnetoexciton luminescence line.

# 1. Introduction

In the present paper we consider the interaction of 2D magnetoexcitons and the possibility of formation a molecule state of the bi-magnetoexciton. The model of the electron-hole (e-h) system consists of the conduction electrons and of the holes in the valence band of the semiconductor layer in a strong perpendicular magnetic field. The particles undergo the Landau quantization (LQ) with the cyclotron energies $\hbar\omega_{ci} = \hbar eB/(m_i c)$, $i = e, h$, where $m_i$ are their effective masses and $B$ is the magnetic field strength. The radii of the cyclotron orbits do not depend on the electron and hole masses $m_i$, being determined only by the magnetic length $l_0 = \sqrt{\hbar c/(eB)}$. The Coulomb e-h interaction gives rise to the magnetoexciton formation, and we assume that electrons and hole occupy the lowest Landau levels (LLLs). The Lorentz force gives rise a strong dependence between the relative and the center-of-mass motions of the e-h pair. As a result the value of ionization potential of the-magnetoexcitons $I_l$, which depends on the wave vector $\vec{k}$, is smaller than the cyclotron energies. The magnetic mass and the dispersion law of the magnetoexciton originate from the Coulomb e-h interaction. For the magnetoexciton with $k = 0$, the electron and hole LQ orbits in Landau gauge description have the same gyration points and the same radii. Therefore, such magnetoexciton looks as a neutral bound particle. For the first time Lerner and Lozovik [1] have shown that magnetexcitons with $k = 0$ does not interact in the LLLs approximation forming an ideal Bose gas, which was confirmed by more detailed studies [2–4]. On the contrary, the magnetoexcitons with $k \neq 0$ have a structure of an electric dipole (shown in Fig.1), since in this case the electron and hole orbits do not overlap. The arm of such dipole moment $d = kl_0^2$ is perpendicular to the wave vector $\vec{k}$.

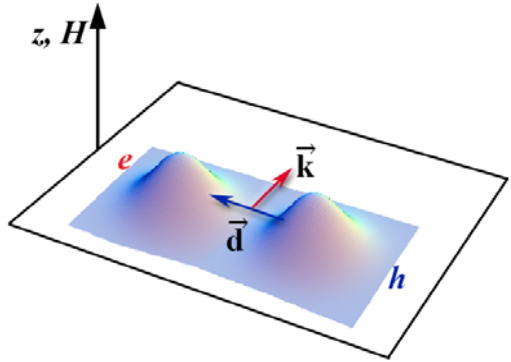

Fig. 1. The electric dipole model of 2D magnetoexcioton with wave vector $\vec{k}$ and the dipole moment $\vec{d}$.

In the present paper we investigate the possibility of a bound state of the magnetic bi-excitons. We consider the bound state of two magnetoexcitons moving with in-plane wave vectors $\vec{k}$ and $-\vec{k}$ forming two electric dipoles with antiparallel dipole moments with the total wave vector equal to zero. The relative exciton-exciton motion is described by the variational Gaussian-type wave functions $\varphi_n(k) \sim (kl_0)^n \exp(-\alpha(kl_0)^2)$ with $n = 0, 2$ depending on the modulus $k$. For $n = 0$ the wave function has the maximum at the point $k = 0$, whereas the other one with $n = 2$ has the maximum on the in-plane ring with the radius $k_r = 1/(\sqrt{\alpha}l_0)$. The geometric structure of the excitonic cloud in the frame of the bound state is an important argument in favor of the choice of Gaussian-type wave functions. Besides, it allows analytical analysis of the problem in question.
The possibility of the formation of bound states essentially depends on mutual orientation of the spins of two electrons and of two holes. We consider the singlet-singlet and triplet-triplet states. In the first case two electrons and two holes are separately in a singlet state with the resultant spin $S = 0$ and its projection on the magnetic field $S_z = 0$. In the second case the total spin of the system is $S = 1$ and $S_z = 0$. We will show that due to the hidden symmetry the stable bound states of the bi-magnetoexcitons do not exist in the LLLs approximation, there is a metastable bound state with considerable energy activation barrier.

The holes taking part in the formation of magnetoexcitons can appear not only in the valence bands of the single or double quantum wells, but also in the case of the two-dimensional electron gas (2DEG) with conduction electrons on the Landau quantization levels with filling factors $v = 1$. The quantum transitions from the filled Landau levels in the next empty Landau levels gives rise to the free electrons in the empty levels and to free holes in the filled levels. This problem was studied by Kallin and Halperin [5] for the conditions of the integer quantum Hall effects (IQHEs) and spinless electrons. Eisenstein and Mac Donald [6] investigated the formation of magnetoexcitons for the case, when electrons are injected in two-layer structure with the filling factor $v = 1/2$. Wojs and coauthors [7–10] have considered the spin waves when excitations take place with the reversing of the spins between the Zeeman splitted levels. In both cases the Landau quantization and Coulomb interactions are the same. In Refs [5–11] the authors pointed on the similarity and even an exact mapping between two-spin and two-charge 2D systems.

Bychkov and Rashba [12] have studied the necessary conditions for the formation of 2D magnetic bi-excitons and they pointed that the asymmetry of the Landau quantization states of the electrons and of the holes is needed. The asymmetry formulated in Ref. [12] means, that some or all matrix elements of the electron-electron (e-e), hole-hole (h-h) and electron-hole (e-h) interactions must coincide to avoid the hidden symmetry of the electrons and of the holes. The e-h asymmetry in the Hamiltonian alters qualitatively the behavior of the bipolar system and can lead to the formation of the magnetic bi-excitons and possibly of the polyexcitons [12]. In this case the scattering amplitude of two magnetoexcitons diverges, when their relative velocity decreases and the formation of the magnetic bi-excitons becomes possible. This condition was obtained for the stable bound states of spinless particles and do not exclude the formation of metastable bound states at some spin structure of two electrons and of two holes. We have shown previously [13] that the influence of the external electric field perpendicular to the layer and the Rashba spin-orbit coupling (RSOC) remove the hidden symmetry leading to the interaction of two magnetic bi-excitons with wave vectors $k=0$, so that making possible bound states of the magnetic bi-excitons. We will show that only a metastable bound state exists, which is of considerable interest since a new luminescence band can appear due to the radiative annihilation of one magnetoexciton taking part in the bound state. But instead of the usual luminescence line arising in the absence of a strong magnetic field due to the bi-exciton-exciton conversion on the energy scale at smaller energies than the exciton-luminescence line, the new luminescence band will be situated at the energies greater than the magnetoexciton luminescence line.

The bi-excitons in $In_xGa_{1-x}As/GaAs$ quantum wells in high magnetic fields were investigated experimentally using the four-wave mixing method [17]. Surprisingly, it was found that the bi-exciton binding energy does not depend on the magnetic field strengths up till 8T, which strength is not sufficient to deal with the magnetic excitons and bi-excitons.

## 2. The Hamiltonian of the electron-hole system and wave functions of the bound states of the interacting 2D magnetoexcitons

In the Landau gauge the charged particles have a free motion in one in-plane direction described by the plane waves with one-dimensional wave numbers $p$ and $q$ and undergo the quantized oscillations around the gyration points in the perpendicular in-plane direction. For electrons and holes at LLLs the quantum numbers of the Landau quantized levels are $n_e = n_h = 0$. The creation and annihilation operators for the electrons and $a^+_{p,\sigma}, a_{p,\sigma}$ and $b^+_{q,\sigma}, b_{q,\sigma}$, correspondingly, have a supplementary spin label $\sigma = \pm 1/2$, which describes the spin projections of the conduction electrons and the effective spin of the heavy holes. A simple generalization of the Hamiltonian describing the Coulomb interaction of 2D electrons and holes situated on their LLLs, neglecting for the sake of simplicity by the electron-hole exchange interaction leading to the formation of the ortho and para magnetoexcitons, as well as by the RSOC, has the form

$$H^{LLL}_{Coul} = \frac{1}{2}\sum_{\vec{Q}} W(\vec{Q})[\hat{\rho}(\vec{Q})\hat{\rho}(-\vec{Q}) - \hat{N}_e - \hat{N}_h], \ \hat{\rho}(\vec{Q}) = \hat{\rho}_e(\vec{Q}) - \hat{\rho}_h(\vec{Q}); W(\vec{Q}) = \frac{2\pi e^2}{\varepsilon_0 S|\vec{Q}|} e^{-\frac{Q^2 l_0^2}{2}},$$

$$\hat{\rho}_e(\vec{Q}) = \sum_{t,\sigma} e^{iQ_y t l_0^2} a^+_{t+\frac{Q_x}{2},\sigma} a_{t-\frac{Q_x}{2},\sigma}; \hat{N}_e = \hat{\rho}_e(0), \ \hat{\rho}_h(\vec{Q}) = \sum_{t,\sigma} e^{-iQ_y t l_0^2} b^+_{t+\frac{Q_x}{2},\sigma} b_{t-\frac{Q_x}{2},\sigma}; \hat{N}_h = \hat{\rho}_h(0), \quad (2)$$

where $\varepsilon_0$ is the dielectric constant, $S$ is the layer surface area, $\hat{\rho}_e(\vec{Q})$ and $\hat{\rho}_h(\vec{Q})$ are the electron and hole plasmon operators, correspondingly.

The Hamiltonian (2) can be transcribed in the way

$$H^{LLL}_{Coul} = H^{LLL}_{e-e} + H^{LLL}_{h-h} + H^{LLL}_{e-h},$$

$$H^{LLL}_{e-e} = \frac{1}{2}\sum_{\vec{Q}}\sum_{p,q}\sum_{\sigma_1,\sigma_2} W(\vec{Q}) e^{-iQ_x Q_y l_0^2} e^{iQ_y(p-q)l_0^2} a^+_{p,\sigma_1} a^+_{q,\sigma_2} a_{q+Q_x,\sigma_2} a_{p-Q_x,\sigma_1},$$

$$H^{LLL}_{h-h} = \frac{1}{2}\sum_{\vec{Q}}\sum_{p,q}\sum_{\sigma_1,\sigma_2} W(\vec{Q}) e^{iQ_x Q_y l_0^2} e^{-iQ_y(p-q)l_0^2} b^+_{p,\sigma_1} b^+_{q,\sigma_2} b_{q+Q_x,\sigma_2} b_{p-Q_x,\sigma_1},$$

$$H_{e-h}^{LLL} = -\sum_{\vec{Q}} \sum_{p,q} \sum_{\sigma_1,\sigma_2} W(\vec{Q}) e^{iQ_y(p+q)l_0^2} a^+_{p,\sigma_1} b^+_{q,\sigma_2} b_{q+Q_x,\sigma_2} a_{p-Q_x,\sigma_1}. \qquad (3)$$

The interaction coefficients depend only on the difference ($p-q$) in for the electron-electron (e-e) and the hole-hole (h-h) interactions, and on the sum ($p+q$) for the electron-hole (e-h) interactions. The magnetoexciton creation operator, which was introduced in [3] and later used in [4, 13] with spin labels, is

$$\hat{\psi}^+_{ex}(\vec{k},\Sigma_e,\Sigma_h) = \frac{1}{\sqrt{N}} \sum_t e^{ik_y t l_0^2} a^+_{t+\frac{k_x}{2},\Sigma_e} b^+_{-t+\frac{k_x}{2},\Sigma_h}, \quad N = \frac{S}{2\pi l_0^2}. \qquad (4)$$

Here $\vec{k}(k_x,k_y)$ is the vector of the center of mass in-plane motion, $t$ is the unidimensional vector of the relative e-h motion with the function of the relative motion $e^{ik_y t l_0^2}$ in the momentum representation, which leads to the $\delta(y-k_y l_0^2)$ function of the relative motion in the real space representation. $N$ is the degree of the degeneracy of the Landau quantization levels, which is proportional to $S$.

The wave function of the magnetoexciton looks as

$$\left|\psi_{ex}\left(\vec{k},\Sigma_e,\Sigma_h\right)\right\rangle = \hat{\psi}^+_{ex}\left(\vec{k},\Sigma_e,\Sigma_h\right)|0\rangle; \; a_{t,\sigma}|0\rangle = b_{t,\sigma}|0\rangle = 0, \qquad (5)$$

where $|0\rangle$ is the ground state of the system. The 2D magnetoexciton with wave vector $\vec{k} \neq 0$ has the form of an electric dipole with the arm $d = kl_0^2$ oriented perpendicularly to the wave vector $\vec{k}$. As it was mentioned above, two magnetoexcitons with wave vectors $\vec{k} = 0$ are similar to neutral compound particles. They have no the dipole moments and do not interact through Coulomb forces. On the contrary, two magnetoexcitons with nonzero wave vectors $\vec{k}_1$ and $\vec{k}_2$ do interact, which opens the opportunity to form a bi-magnetoexciton. The wave function of two magnetoexcitons with quantum numbers $\left|\vec{k},\Sigma_{e1},\Sigma_{h1}\right\rangle$ and $\left|-\vec{k},\Sigma_{e2},\Sigma_{h2}\right\rangle$ is

$$\left|\psi_{ex,ex}\left(\vec{k},\Sigma_{e1},\Sigma_{h1};-\vec{k},\Sigma_{e2},\Sigma_{h2}\right)\right\rangle = \frac{1}{N} \sum_{t,s} e^{ik_y(t-s)l_0^2} a^+_{t+\frac{k_x}{2},\Sigma_{e1}} a^+_{s-\frac{k_x}{2},\Sigma_{e2}} b^+_{-s-\frac{k_x}{2},\Sigma_{h2}} b^+_{-t+\frac{k_x}{2},\Sigma_{h1}} |0\rangle. \qquad (6)$$

The wave function of the quasi bi-magnetoexciton with wave vector $\vec{k}=0$ as a bound state of two magnetoexcitons with wave vectors $\vec{k}$ and $-\vec{k}$ and spin quantum numbers $\Sigma_{e1}, \Sigma_{h1}, \Sigma_{e2}, \Sigma_{h2}$ can be constructed as a superposition of the wave functions (6) introducing the wave function $\varphi_n(\vec{k})$ of the relative motion, which ~~can~~ play also the role of the variational function determining the minimal energy of the bimagnetoexciton, as well as the density $|\varphi_n(\vec{k})|^2$ of the magnetoexcitons taking part in the formation of the bound state. The spin configurations of the bound states depend essentially on the ratio between the ortho-para exciton splitting and the binding energy of the bi-exciton. In the presence of a strong magnetic field these values are unknown and we will be determined below. For this purpose we construct the symmetric and antisymmetric superpositions of two electron spin states and two hole effective spin states in the form

$$\frac{1}{\sqrt{2}} \sum_{\Sigma_e = \pm 1/2} (\eta_e)^{\Sigma_e + \frac{1}{2}} a^+_{p,\Sigma_e} a^+_{q,-\Sigma_e}; \quad \frac{1}{\sqrt{2}} \sum_{\Sigma_h = \pm 1/2} (\eta_h)^{\Sigma_h + \frac{1}{2}} b^+_{p,\Sigma_h} b^+_{q,-\Sigma_h}; \quad \eta_e = \pm 1; \quad \eta_h = \pm 1; \quad (7)$$

Here we took into account that the electron-hole exchange interaction is neglected in the Hamiltonian (2) and (3). Below we will suppose that both pairs of spins $(\Sigma_{e1}, \Sigma_{e2})$ and $(\Sigma_{h1}, \Sigma_{h2})$ are simultaneously in the states with the same $\eta_e = \eta_h = \eta = \pm 1$. The wave functions of the bi-magnetoexcitons for these conditions are

$$\left| \psi_{bimex}(0, \eta, \varphi_n) \right\rangle = \frac{1}{2N^{3/2}} \sum_{\Sigma_e, \Sigma_h} (\eta)^{\Sigma_e + \Sigma_h + 1} \sum_{\vec{k}} \varphi_n(\vec{k}) \sum_{s,t} e^{ik_y(t-s)l_0^2} a^+_{t+\frac{k_x}{2},\Sigma_e} a^+_{s-\frac{k_x}{2},-\Sigma_e} b^+_{-s-\frac{k_x}{2},-\Sigma_h} b^+_{-t+\frac{k_x}{2},\Sigma_h} |0\rangle. \quad (8)$$

The chosen variational wave functions of the relative motion in the momentum and in the real space representations $\varphi_n(\vec{k})$ and $\psi_n(\vec{r})$, their normalization conditions and the main parameters are

$$\varphi_0(x) = (4\alpha)^{1/2} e^{-\alpha x^2}; \quad \varphi_2(x) = (8\alpha^3)^{1/2} x^2 e^{-\alpha x^2}; \quad x = kl_0, \quad \frac{1}{N} \sum_{\vec{k}} |\varphi_n(\vec{k})|^2 = \int_0^\infty x dx |\varphi_n(x)|^2 = 1,$$

$$\psi_n(\vec{r}) = \int \varphi_n(\vec{k}) e^{i\vec{k}\vec{r}} d^2\vec{k} = \int_0^\infty x dx \varphi_n(x) J_0(x \cdot r/l_0), \quad \psi_0(r) \sim e^{-\frac{r^2}{4\alpha l_0^2}}; \quad \psi_2(r) \sim \left(1 - \frac{r^2}{4\alpha l_0^2}\right) e^{-\frac{r^2}{4\alpha l_0^2}}. \quad (9)$$

Here $J_0(z)$ is the Bessel function of the zeroth order. The selected variational wave functions depend only on the modulus $k = |\vec{k}|$. $\varphi_0(x)$ has a maximum at the point $x = 0$, the mean value $\overline{x^2} = 1/(2\alpha)$, the radius of the quantum state $\psi_0(\vec{r})$ equals to $a = 2l_0\sqrt{\alpha}$. The function $\varphi_2(\vec{k})$ has the maximum on the 2D ring with the radius $k_r = 1/(l_0\sqrt{\alpha})$. The magnetoexciton densities for the bound states $|\varphi_2(\vec{k})|^2$ and $|\varphi_0(\vec{k})|^2$ are shown in Fig. 2. In the real space the function $\psi_2(r)$ has a maximum at the point $r_0 = 0$, which is positive for $r_1 \leq a$, where $\psi_2(r)$ changes sign and reaches a minimum at the point $r_2 = l_0\sqrt{8\alpha}$. In fact the function $\psi_2(r)$ has the same radius $a$, the absolute value at the minimum is much smaller than at the maximum.

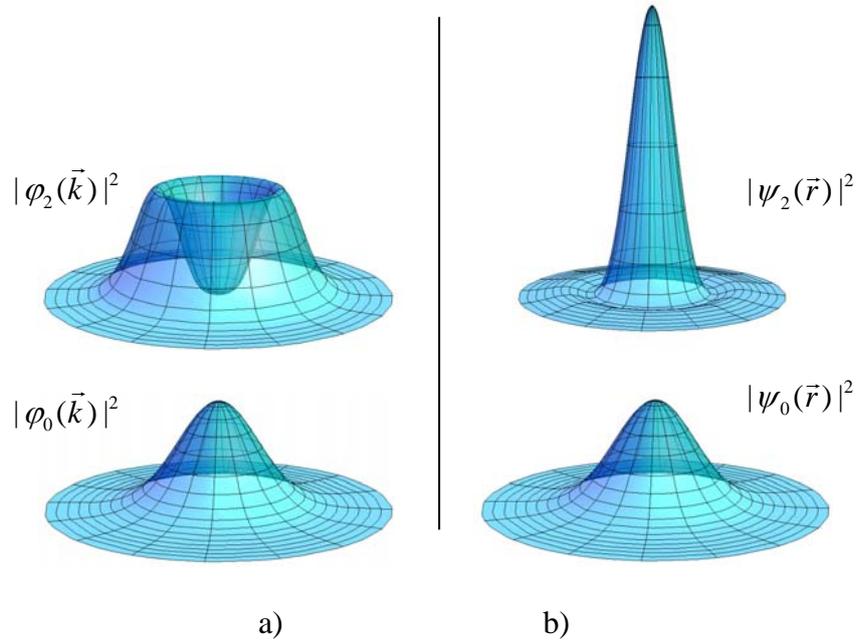

a) b)

Fig. 2. The magnetoexciton densities in the frame of the bound states: a) $|\varphi_2(\vec{k})|^2$ and $|\varphi_0(\vec{k})|^2$ in the momentum space representation, b) $|\psi_2(r)|^2$ and $|\psi_0(r)|^2$ in the real space representation.

Contrary to the function $\varphi_0(x)$, the normalization integral of the quasi bi-magnetoexciton wave function (8) calculated for the function $\varphi_2(x)$, give rise to the overlapping integrals $L_n(\alpha)$ as follows

$$\langle \phi_{bimex}(0,\eta,\varphi_n) | \phi_{bimex}(0,\eta,\varphi_n) \rangle = 2(1 - \eta L_n(\alpha)),$$

$$L_n(\alpha) = \int_0^\infty xdx \int_0^\infty ydy \varphi_n^*(x)\varphi_n(y)J_0(xy); \quad L_0(\alpha) = \frac{1}{\alpha + 1/(4\alpha)}; \quad L_2(\alpha) = \frac{2\alpha^2 - 1/\alpha}{(\alpha + 1/(4\alpha))^3}. \tag{10}$$

### 3. Binding energies of the bound states of the two interacting magnetoexcitons

The expectation values of the Hamiltonian (3) averaged over the wave functions (8), which is characterized by the wave vector $\vec{k} = 0$, by two values of $\eta = \pm 1$, and by the variational wave functions $\varphi_n(\vec{k})$, equal to

$$E_{bimex}(0,\eta,\varphi_n) = \frac{\langle \phi_{bimex}(0,\eta,\varphi_n) | H_{Coul}^{LLL} | \phi_{bimex}(0,\eta,\varphi_n) \rangle}{\langle \phi_{bimex}(0,\eta,\varphi_n) | \phi_{bimex}(0,\eta,\varphi_n) \rangle}. \tag{11}$$

Integration over the angle variables and excluding the trial function $\varphi_n(|\vec{x}-\vec{y}|)$ leads to the expression

$$\langle \psi_{bimex}(0,\eta,\varphi_n) | H_{coul}^{LLL} | \psi_{bimex}(0,\eta,\varphi_n) \rangle = -4\left(\frac{e^2}{\varepsilon_0 l_0}\right) \int_0^\infty dy\, e^{-\frac{y^2}{2}} \times$$

$$\times \int_0^\infty xdx |\varphi_n(x)|^2 J_0(xy) + 4\left(\frac{e^2}{\varepsilon_0 l_0}\right) \int_0^\infty dy\, e^{-\frac{y^2}{2}} \int_0^\infty xdx \varphi_n^*(x) \frac{1}{2\pi} \int_0^{2\pi} d\theta \times$$

$$\times \frac{1}{2\pi} \int_0^{2\pi} d\psi \varphi_n(|\vec{x}-\vec{y}|)\left(e^{ix \cdot y \sin(\theta-\psi)} - 1\right) + 4\eta\left(\frac{e^2}{\varepsilon_0 l_0}\right) \int_0^\infty dy\, e^{-\frac{y^2}{2}} \times$$

$$\times \int_0^\infty xdx \varphi_n^*(x) \int_0^\infty zdz \varphi_n(z)\left(J_0(x \cdot y)J_0(x \cdot z) + J_0(x \cdot z)J_0(y \cdot z) - \right.$$

$$\left. -J_0(x \cdot y)J_0(x \cdot z)J_0(y \cdot z) - 2\sum_{k=1}^\infty J_{2k}(x \cdot y)J_{2k}(x \cdot z)J_{2k}(y \cdot z)\right), \tag{12}$$

where $|\vec{x}-\vec{y}| = \sqrt{x^2 + y^2 - 2xy\cos(\theta-\psi)}$, $\vec{x} = (x\cos\theta, x\sin\theta)$, $\vec{y} = (y\cos\psi, y\sin\psi)$,

Here the variational wave function $\varphi_2(x)$ was taken in the form

$$\varphi_2(|\vec{x}-\vec{y}|) = (8\alpha^3)^{1/2} |\vec{x}-\vec{y}|^2 e^{-\alpha|\vec{x}-\vec{y}|^2} = (8\alpha^3)^{1/2} e^{-\alpha(x^2+y^2)} \times$$

$$\times \left[x^2 + y^2 - 2xy\frac{\partial}{\partial(2\alpha xy)}\right]\left[I_0(2\alpha xy) + 2\sum_{k=1}^\infty I_k(2\alpha xy)\cos(k(\theta-\psi))\right], \tag{13}$$

and the derivatives of the Bessel functions $I_n(z)$ and $J_n(z)$ of the integer order were used [20–22].

The second term in the right hand side of the average value (12) can be transcribed in the following way

$$\varepsilon_2(\varphi_2,\eta,\alpha) = 4\left(\frac{e^2}{\varepsilon_0 l_0}\right)\int_0^\infty dy e^{-\frac{y^2}{2}}\int_0^\infty x dx \varphi_2^*(x)\frac{1}{2\pi}\int_0^{2\pi}d\theta\frac{1}{2\pi}\int_0^{2\pi}d\psi \times$$

$$\times \varphi_2(|\vec{x}-\vec{y}|)(e^{ixy\sin(\theta-\psi)}-1) = 4\left(\frac{e^2}{\varepsilon_0 l_0}\right)(8\alpha^3)\times$$

$$\times\left\{\int_0^\infty dy e^{-y^2\left(\frac{1}{2}+\alpha\right)}\int_0^\infty dx x^5 e^{-2\alpha x^2}J_0(x\cdot y)I_0(2\alpha xy) + \int_0^\infty dy y^2 e^{-y^2\left(\frac{1}{2}+\alpha\right)}\int_0^\infty dx x^3 e^{-2\alpha x^2}J_0(xy)I_0(2\alpha xy) - \right.$$

$$-\int_0^\infty dy e^{-y^2\left(\frac{1}{2}+\alpha\right)}\int_0^\infty dx x^5 e^{-2\alpha x^2}I_0(2\alpha xy) - 2\int_0^\infty dy\cdot y e^{-y^2\left(\frac{1}{2}+\alpha\right)}\int_0^\infty dx x^4 e^{-2\alpha x^2}J_0(xy)I_1(2\alpha xy) -$$

$$-\int_0^\infty dy y^2 e^{-y^2\left(\frac{1}{2}+\alpha\right)}\int_0^\infty dx x^3 e^{-2\alpha x^2}I_0(2\alpha xy) + 2\int_0^\infty dy\cdot y e^{-y^2\left(\frac{1}{2}+\alpha\right)}\int_0^\infty dx x^4 e^{-2\alpha x^2}I_1(2\alpha xy) +$$

$$+2\int_0^\infty dy e^{-y^2\left(\frac{1}{2}+\alpha\right)}\int_0^\infty dx\cdot x^5 e^{-2\alpha x^2}\sum_{k=1}^\infty J_{2k}(xy)I_{2k}(2\alpha xy) -$$

$$-4\int_0^\infty dy\cdot y e^{-y^2\left(\frac{1}{2}+\alpha\right)}\int_0^\infty dx x^4 e^{-2\alpha x^2}\sum_{k=1}^\infty J_{2k}(xy)I_{2k+1}(2\alpha xy) +$$

$$+2\int_0^\infty dy y^2 e^{-y^2\left(\frac{1}{2}+\alpha\right)}\int_0^\infty dx x^3 e^{-2\alpha x^2}\sum_{k=1}^\infty J_{2k}(xy)I_{2k}(2\alpha xy) -$$

$$\left. -\frac{4}{\alpha}\int_0^\infty dy e^{-y^2\left(\frac{1}{2}+\alpha\right)}\int_0^\infty dx x^3 e^{-2\alpha x^2}\sum_{k=1}^\infty k J_{2k}(xy)I_{2k}(2\alpha xy)\right\}. \qquad (14)$$

The third term in the right hand side of (12) can be written as

$$\varepsilon_3(\varphi_2,\eta,\alpha) = 4\eta\left(\frac{e^2}{\varepsilon_0 l_0}\right)\int_0^\infty dy e^{-\frac{y^2}{2}}\int_0^\infty x dx \varphi_2^*(x)\int_0^\infty z dz \varphi_2(z)\left(J_0(x\cdot y)J_0(x\cdot z) + J_0(x\cdot z)J_0(y\cdot z) - \right.$$

$$-J_0(x\cdot y)J_0(x\cdot z)J_0(y\cdot z) - 2\sum_{k=1}^\infty J_{2k}(x\cdot y)J_{2k}(x\cdot z)J_{2k}(y\cdot z)\bigg) =$$

$$= 4\eta\left(\frac{e^2}{\varepsilon_0 l_0}\right)(8\alpha^3)\int_0^\infty dy e^{-\frac{y^2}{2}}\int_0^\infty dx x^3 e^{-\alpha x^2}\int_0^\infty dz z^3 e^{-\alpha z^2}\left[J_0(x\cdot y)J_0(x\cdot z) + J_0(x\cdot z)J_0(y\cdot z) - \right.$$

$$\left.-J_0(x\cdot y)J_0(x\cdot z)J_0(y\cdot z) - 2\sum_{k=1}^\infty J_{2k}(x\cdot y)J_{2k}(x\cdot z)J_{2k}(y\cdot z)\right]. \qquad (15)$$

The first term in the average value (12), and the overlapping integrals $L_n(\alpha)$ can be calculated analytically.

$$\varepsilon_1(\varphi_2,\eta,\alpha) = -4\left(\frac{e^2}{\varepsilon_0 l_0}\right)\int_0^\infty dy\, e^{-\frac{y^2}{2}} \int_0^\infty dx\, x|\varphi_2(x)|^2 J_0(xy) = -4I_l \cdot \frac{\sqrt{4\alpha}}{\sqrt{1+4\alpha}}\left[1 - \frac{1}{1+4\alpha} + \frac{3}{8}\frac{1}{(1+4\alpha)^2}\right];$$

$$I_l = \left(\frac{e^2}{\varepsilon_0 l_0}\right)\sqrt{\frac{\pi}{2}};\quad L_2(\alpha) = \frac{2\alpha^2 - \frac{1}{2}}{\left(\alpha + \frac{1}{4\alpha}\right)^3}. \tag{16}$$

Here $I_l$ is the ionization potential of the 2D magnetoexciton with wave vector $\vec{k}_\parallel = 0$.

$$\frac{E_1(\varphi_2,\eta,\alpha)}{2I_l} = \frac{\varepsilon_1(\varphi_2,\eta,\alpha)}{4I_l(1-\eta L_2(\alpha))} = -\frac{\sqrt{4\alpha}\left[1 - \frac{1}{1+4\alpha} + \frac{3}{8}\frac{1}{(1+4\alpha)^2}\right]}{\sqrt{1+4\alpha}\left[1 - \frac{\eta(2\alpha^2 - 1/2)}{(\alpha + 1/(4\alpha))^3}\right]}. \tag{17}$$

The expressions (14)-(16), as well as the denominator (10) contain the integrals with one, two and three Bessel functions. The integrals $I_1 - I_{13}$ can be calculated analytically and the corresponding calculations are presented in Appendix A. Figure 3 shows the dependence on the parameter $\alpha$ of the variational wave function $\varphi_2(x)$.

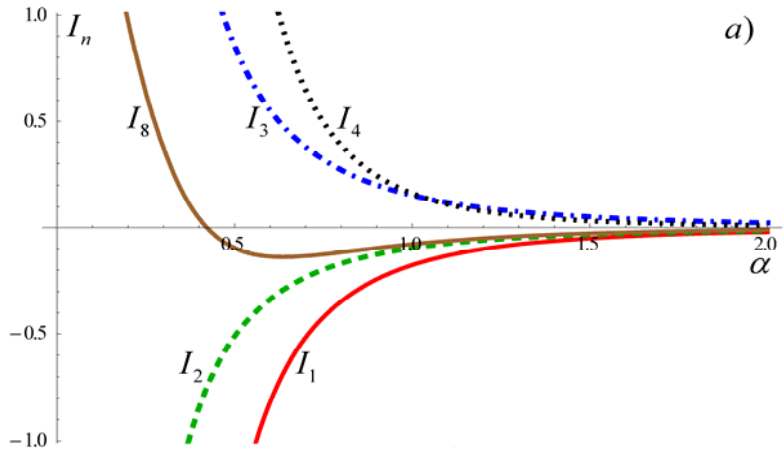

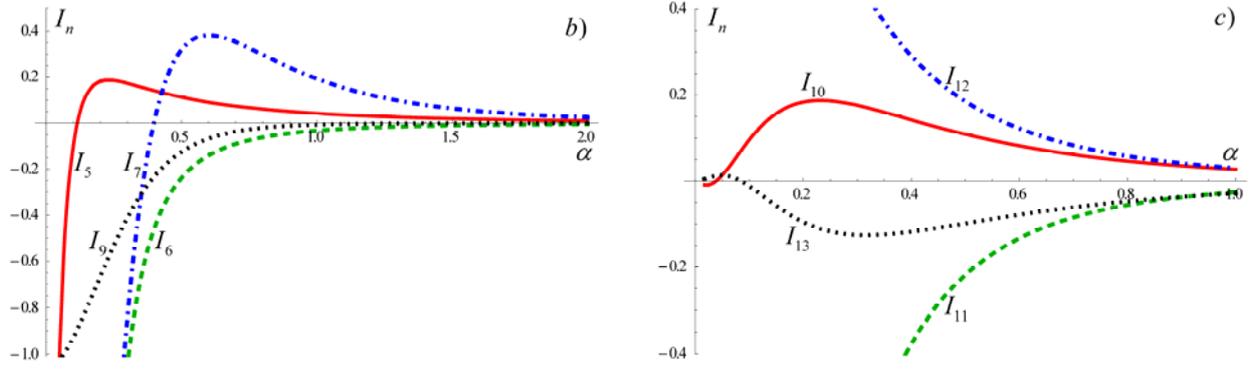

Fig. 3. The integrals $I_1 - I_{13}$ in dependence on the parameter $\alpha$ of the variational wave function $\varphi_2(k)$ are represented in the three groups: a), b) and c).

Since the denominator has small values at some parameter $\alpha$, the analytical calculations are helpful. In this case the small deviation of the numerator in the expression (11) from its exact values could lead to wrong conclusions concerning the bound states.

## 4. The electron structure of the bound states

The most interesting results concern the function $\varphi_2(\vec{k})$. It describes the electron structure of the bound state, when the maximal density of the magnetoexciton cloud in the momentum space representation is concentrated on the in-plane ring with the radius $k_r = 1/(\sqrt{\alpha} l_0)$, presented in the Fig. 2. In the real space representation the exciton cloud has maximum density at the center of the structure with the radius of the dome $R \sim 2\sqrt{\alpha} l_0$ depending on the parameter $\alpha$. The results are completely different for two spin configurations with $\eta = \pm 1$. In both spin configurations the full energies of the bound states are greater than the value $-2I_l$, where $I_l$ is the ionization potential of the free 2D magnetoexciton with $\vec{k} = 0$. All these bound states are unstable as regards the dissociation in the form of two free magnetoexcitons with $\vec{k} = 0$. However, in the case $\eta = 1$ and $\alpha = 0.5$, a deep metastable bound state with considerably large activation barrier comparable with the magnetoexciton ionization potential $I_l$ was revealed. On the contrary, in the case $\eta = -1$ and $\alpha = 3.4$ only a shallow bound state was found with nonsignificant barrier. Fig. 4 shows the total energies of two bound 2D magnetoexcitons with wave vectors $\vec{k}$ and $-\vec{k}$, for different spin structures $\eta = \pm 1$.

The obtained results with the variational function $\varphi_0(\vec{k})$ clearly demonstrate that the magnetoexcitons with the maximal density in the point $\vec{k} = 0$ of the momentum space practically do not interact due to their hidden symmetry. The energies of two magnetoexcitons are very close to the value $-2I_l$ at any values of the parameter $\alpha$ with the exception of the case $\eta = 1$ and $\alpha = 0.5$, where the singularity appeared due to the zero value of the denominator.

The obtained results can be better understood taking into account the dipole-dipole interactions in 2D e-h systems formulated by Wojs and coauthors in the Refs [7–10] in the case of two spin waves, as well as by Olivares-Robles and Ulloa [11] in the case of magnetoexcitons with spatially separated electrons and holes. In the case of two spin waves discussed in [7], the numerical diagonalization method in the Haldane geometry [14] was used. It was shown that two spin waves moving in-plane in the same direction with parallel dipole moments attract each other, which leads to their binding. The two spin waves moving in opposite directions with antiparallel electric dipoles undergo the repulsion [7]. The magnetoexcitons with electrons and holes spatially separated in two wells of the double quantum well with particles moving in parallel planes are characterized supplementary to the in-plane dipoles by the static dipole moments oriented perpendicularly to the layers. They give rise to the preponderant repulsion between these magnetoexcitons. Nevertheless two magnetoexcitons moving with the parallel wave vectors and with parallel in-plane dipole moments have a total less repulsive interaction. In our case we have deal with the interaction of two magnetoexcitons with antiparallel wave vectors $\vec{k}$ and $-\vec{k}$, and antiparallel electric dipole moments, that have repulsive interaction. This is correct when the mean distance $R \simeq 2\sqrt{\alpha}l_0$ between two magnetoexcitons in the bound state is much greater than the arm $d = kl_0^2 \approx l_0/\sqrt{\alpha}$ of the dipole moment. The condition $R >> d$ means $\alpha >> 1/2$. In this range of the parameter $\alpha$ the repulsion between two magnetoexcitons take place and their bound state is unstable. The antiparallel dipoles at $\alpha \gg 1/2$ prevent the formation of the bound states of two 2D magnetoexcitons. As was mentioned above the 2D magnetoexcitons with $\vec{k} \neq 0$ look like electric dipoles with in-plane arms having the length $d = kl_0^2$, which are perpendicular to the direction of the wave vectors $\vec{k}$. The bound (molecule) states can be formed by two magnetoexcitons with antiparallel wave vectors $\vec{k}$ and $-\vec{k}$. They have the structure of two antiparallel dipoles bound together. Their possible orientation as a whole in any direction of the layer plane with equal probability was supposed. Such possibility corresponds to introducing the trial wave function of the relative motion of two magnetoexcitons in the frame of the bound state $\varphi_n(\vec{k})$, which depends on the modulus $k$.

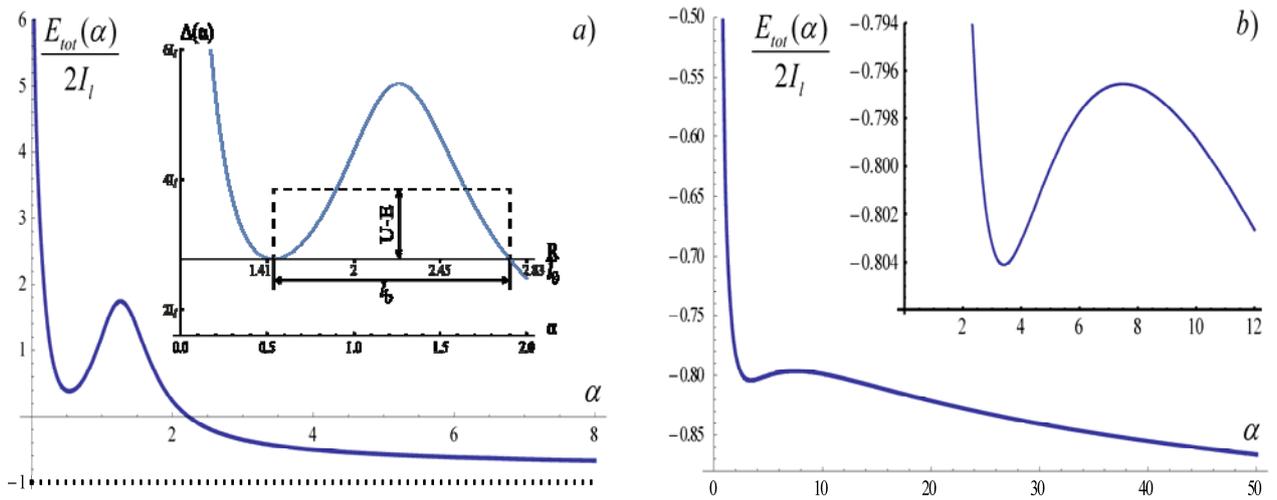

Fig. 4. The total energies of two bound 2D magnetoexcitons with wave vectors $\vec{k}$ and $-\vec{k}$, with different spin structures $\eta = \pm 1$ and with the variational wave function $\varphi_2(k)$, in dependence on the parameter $\alpha$. a) the case $\eta = 1$, b) the case $\eta = -1$. The total energies are normalized to the value $2I_l$, where $I_l$ is the ionization potential of a free magnetoexciton with wave vector $\vec{k} = 0$.

In both spin configurations $\eta = \pm 1$, the full energies of the bound states are greater than the value $-2I_l$ in all range of the values $\alpha$, as is shown in Fig. 2. All these states are unstable in respect to dissociation into two free magnetoexcitons with $\vec{k} = 0$. There are a deep metastable bound state with the activation barrier comparable with two magnetoexciton ionization potentials $2I_l$ in the case $\eta = 1$ and $\alpha = 0.5$, and a weakly bound state in the case $\eta = -1$ and $\alpha = 3.4$. The metastable bound state with $\alpha = 0.5$ and $\eta = 1$ can give rise to a new luminescence band due to its radiative conversion into the para magnetoexciton state with resultant spin of the electron-hole pair and its projection on the magnetic field equal to zero. Calculations [20] of the corresponding matrix elements have demonstrated that the transition in the para magnetoexciton state is allowed, whereas in the ortho magnetoexciton states with $S = 1$ and $S_z = 0, \pm 1$ the transitions are forbidden.

The position of the new luminescence band on the energy scale is situated on the high energy side relative to the para magnetoexciton luminescence line. The shift $\Delta(\alpha)$ equals to $\Delta(\alpha) = E_{bimex}(0,\eta,\varphi_n) + 2I_l$ and is shown in the insert in Fig. 4, where the rectangular activation barrier is shown in dependence on the mean distance $R = 2\sqrt{\alpha}l_0$ between two magnetoexcitons in the frame of the bound state. It can be considered as the effective rectangular barrier with the relative height $U - E = 1.35I_l$, which equals to one half of the maximal relative height. The length of the effective rectangular barrier equals to $l_b = 1.34l_0$. This occurs when the emitted photon propagates perpendicularly to the layer and the appeared para magnetoexciton has the in-plane wave vector $Q_\parallel = 0$. In more general case the shift depends on the dispersion law of the para magnetoexciton in the point $\vec{Q}_\parallel \neq 0$.

Let us calculate the tunneling transparency of the effective rectangular barrier for the hypothetical particle with the effective mass $\mu$ equal to one half of the magnetic mass $M(B)$, as it takes place in the relative motion of two magnetoexcitons in the bound state. Taking into account the magnetic mass $M(B) = \hbar^2/(I_l l_0^2)$, we obtain that the coefficient of the transparency $T = \exp[-2l_b\sqrt{2\mu(U-E)}/\hbar] \approx \exp[-3.2] \approx 10^{-1.4} \approx 0.04$ does not depend on the magnetic field strength $B$.

Because the particle is in the metastable state with the parameters $\alpha = 0.5$ and $\eta = 1$ on the back of the effective barrier, in the confined space with the radius $R = 2\sqrt{\alpha}l_0$, it is moving in the effective trap with the velocity $v = \hbar/(\mu R)$ and makes $N_{bl}$ blows per second on the inner side of the effective barrier trying to tunnel from it. Following the theory of the $\alpha$-decay (see, e.g. [21]), we can determine the number of the blows per second as well as the probability $P_{tun}$ of tunneling through the effective rectangular barrier, which is equal to the product of $N_{bl}$ and $T$. After simple quantum mechanical calculations, we obtain:

$$N_{bl} = \frac{v}{2R} = \frac{\hbar}{2\mu R^2} = \frac{\hbar}{4\alpha l_0^2 M(B)} = \frac{I_l}{4\alpha\hbar} \approx \frac{1}{l_0} \; ; \; P_{tun} = N_{bl}T \approx \frac{1}{l_0}; \; \tau = P_{tun}^{-1} \approx l_0 = \sqrt{\hbar c/eB}. \quad (18)$$

In the case of $\alpha = 0,5$ and $l_0 = 10^{-6}$ cm, which corresponds to magnetic field strength $B = 6\,\text{T}$, we can estimate the number $N_{bl} \approx 10^{13}\,\text{sec}^{-1}$ and the tunneling probability $P_{tun} \approx 10^{11.6}\,\text{sec}^{-1}$. This means that the lifetime ($\tau = P_{tun}^{-1}$) of the particle in these conditions equals to 2.5 ps. Both values $N_{bl}$ and $P_{tun}$ increase with the increase of the magnetic field strength, whereas the lifetime $\tau$ decreases.

## 5. Conclusions

The bound states of the 2D bi-magnetoexcitons formed by two magnetoexcitons with opposite wave vectors and with antiparallel electric dipole moments in the singlet-singlet and in the triplet-triplet spin structures were investigated in the lowest Landau levels approximation. It is shown that the stable bound states do not exist due to the hidden symmetry of the electron-hole system. At the same time a metastable bound state in the triplet-triplet spin configuration with an effective energy activation barrier comparable with $2I_l$ was revealed. The coefficient of the tunneling transparency $T$ of the effective barrier was estimated for the hypothetical particle with the mass $\mu$ equal to one half of the magnetoexciton magnetic mass $M(B)$. It happens to be of the order $T \approx \exp[-3.2] \approx 10^{-1.4} \approx 0.04$ and does not depend on the magnetic field strength $B$. The lifetime of the particle in the metastable state depends essentially on $B$ and can be estimated as $\tau = 2.5$ ps at $l_0 = 10^{-6}$ cm and $B = 6\,\text{T}$. The metastable bound state can give rise to a new luminescence band due to the radiative decay and to conversion in the para magnetoexciton with resultant spin of the electron-hole pair $S = 0$ and $S_z = 0$. The new band is situated on the energy scale at the high energy side of the para magnetoexciton luminescence line with the shift $\Delta(\alpha) = E_{bimex}(0, \eta, \varphi_n) + 2I_l$.

It is interesting to note a certain similarity in the formation of the metastable bound state of 2D bi-magnetoexciton with a hydrogen molecule in strong magnetic fields. Liberman and Kravchenko [21, 22, 23] have shown that in a strong magnetic field, in the range $\gamma = 1.2 - 1.4$, where $\gamma = B/B_c$, $B_0 = m_e^2 e^3 c / \hbar^3$) the triplet state $^3\Pi_u$ forms a metastable state of the system, because its minimum lies below the potential curve of the ground state $^3\Sigma_u^+$. Extending the developed theory for the hydrogen-like excitons they suggested [24] that the triplet metastable state $^3\Pi_u$ can be associated with the alternative excitonic bound state and may explain appearance of "X-line" in the optical spectra of a stressed Ge crystal, observed experimentally in [25] at the magnetic field exceeding 4T (for the stressed Ge crystal $B_0$=2.9T).

## Appendix A

The expression (14) includes three integrals containing modified Bessel function $I_\nu(2\alpha xy)$ with $\nu = 0, 1$. They depend on the parameter $\alpha$ of the variational wave function $\varphi_2(x)$ in the way

$$I_1 = -\int_0^\infty dy\, e^{-y^2\left(\frac{1}{2}+\alpha\right)} \int_0^\infty dx\, x^5 e^{-2\alpha x^2} I_0(2\alpha xy) = -\frac{\sqrt{2\pi}}{128\alpha^3} \cdot \frac{\left(8 + 24\alpha + 19\alpha^2\right)}{(1+\alpha)^{5/2}},$$

$$I_2 = -\int_0^\infty dy\, y^2 e^{-y^2\left(\frac{1}{2}+\alpha\right)} \int_0^\infty dx\, x^3 e^{-2\alpha x^2} I_0(2\alpha xy) = -\frac{\sqrt{2\pi}}{32\alpha^2} \cdot \frac{(2+5\alpha)}{(1+\alpha)^{5/2}}, \quad (A.1)$$

$$I_3 = 2\int_0^\infty dy \cdot y \cdot e^{-y^2\left(\frac{1}{2}+\alpha\right)} I_1(2\alpha xy) = \frac{\sqrt{2\pi}}{32\alpha^2} \cdot \frac{(4+7\alpha)}{(1+\alpha)^{5/2}}.$$

Contrary to the first three integrals $I_1$, $I_2$, $I_3$ containing one modified Bessel function $I_0(cx)$ or $I_1(cx)$ another three integrals contain the products of two Bessel functions of the types $J_0(bx)I_0(cx)$ and $J_0(bx)I_1(cx)$, one of them being also the modified Bessel function. The integrals $I_4, I_5$ and $I_6$ can be calculated analytically, using the handbooks [26–27]. The result of the analytical calculations is:

$$I_4 = \int_0^\infty dy\, e^{-y^2\left(\frac{1}{2}+\alpha\right)} \int_0^\infty dx\, x^5 e^{-2\alpha x^2} J_0(xy) I_0(2\alpha xy) = \frac{1}{32\alpha^4}\left\{3I_0^{1/2}(q,c) + \frac{3}{8\alpha}(4\alpha^2-1) I_0^{3/2}(q,c) + \right.$$

$$\left. + \frac{(4\alpha^2-1)}{128\alpha^2} I_0^{5/2}(q,c) - \frac{7}{4} I_1^{3/2}(q,c) - \frac{(4\alpha^2-1)}{16\alpha} I_1^{5/2}(q,c) + \frac{1}{8} I_2^{5/2}(q,c)\right\},$$

$$I_5 = \int_0^\infty dy\, y^2 e^{-y^2\left(\frac{1}{2}+\alpha\right)} \int_0^\infty dx\, x^3 e^{-2\alpha x^2} J_0(xy) I_0(2\alpha xy) = \frac{1}{16\alpha^2}\left\{I_0^{3/2}(q,c) + \frac{(4\alpha^2-1)}{8\alpha} I_0^{5/2}(q,c) - \frac{1}{2} I_1^{5/2}(q,c)\right\},$$

$$I_6 = -2\int_0^\infty dy\, y e^{-y^2\left(\frac{1}{2}+\alpha\right)} \int_0^\infty dx\, x^4 e^{-2\alpha x^2} J_0(xy) I_1(2\alpha xy) =$$

$$= -\frac{1}{16\alpha^3}\left\{I_0^1(q,c) + \frac{(4\alpha^2-1)}{16\alpha} I_0^2(q,c) - \frac{1}{4} I_1^2(q,c) - \frac{3}{2} I_1^{3/2}(q,c) - \frac{(4\alpha^2-1)}{16\alpha} I_1^{5/2}(q,c) + \frac{1}{4} I_2^{5/2}(q,c)\right\}.$$

(A.2)

The third contribution, $\varepsilon_3(\varphi_2,\eta,\alpha)$, described by Eq. (15), is determined by integrals $I_7, I_8$ and $I_9$, where the integral $I_7$ is

$$I_7 = \int_0^\infty dy\, e^{-\frac{y^2}{2}} \int_0^\infty dx\, x^3 e^{-\alpha x^2} \int_0^\infty dz\, z^3 e^{-\alpha z^2} \left(J_0(xy) J_0(xz) + J_0(xz) J_0(yz)\right) =$$

$$= \frac{4\sqrt{\pi}(4\alpha^2-1)}{(4\alpha^2+1)^3 q^{1/2}} - \frac{2\alpha\sqrt{\pi}(4\alpha^2-3)}{(4\alpha^2+1)^4 q^{3/2}} - \frac{3\alpha^2\sqrt{\pi}}{4(4\alpha^2+1)^4 q^{5/2}}.$$

(A.3)

It was calculated exactly taking into account that the product of two Bessel functions $J_0(xz) J_0(yz)$ can be transformed into the expression $J_0(xy) J_0(xz)$ by the interchange of the variables $x \rightleftarrows z$.

Two integrals $I_8$ and $I_9$ contain three Bessel functions

$$I_8 = -\int_0^\infty dy\, e^{-\frac{y^2}{2}} \int_0^\infty dx\, x^3 e^{-\alpha x^2} \int_0^\infty dz\, z^3 e^{-\alpha z^2} J_0(xy) J_0(xz) J_0(yz),$$ (A.4)

and

$$I_9 = -2\sum_{n=1}^\infty \int_0^\infty dy\, e^{-\frac{y^2}{2}} \int_0^\infty dx\, x^3 e^{-\alpha x^2} \int_0^\infty dz\, z^3 e^{-\alpha z^2} \cdot J_{2n}(xy) J_{2n}(xz) J_{2n}(yz).$$ (A.5)

The analytical calculation of integrals (A.4) and (A.5), using the handbooks [26–28], leads to cumbersome expressions, which have been published in [18].

There are still four double integrals $I_{10} - I_{13}$ in the expression (14). They were calculated analytically using the handbooks [26–27] as follows:

$$I_{10} = 2\sum_{n=1}^{\infty}\int_{0}^{\infty} dy\, y^2 e^{-y^2\left(\frac{1}{2}+\alpha\right)}\int_{0}^{\infty} dx\, x^3 e^{-2\alpha x^2} J_{2n}(xy) I_{2n}(2\alpha xy) =$$

$$= \frac{1}{8\alpha^2}\sum_{n=1}^{\infty}\left[(2n+1) I_{2n}^{3/2}(q,c) + \frac{(4\alpha^2-1)}{8\alpha} I_{2n}^{5/2}(q,c) - \frac{1}{2} I_{2n+1}^{5/2}(q,c)\right],$$

$$I_{11} = -\frac{4}{\alpha}\sum_{n=1}^{\infty} n\int_{0}^{\infty} dy\, e^{-y^2\left(\frac{1}{2}+\alpha\right)}\int_{0}^{\infty} dx\, x^3 e^{-2\alpha x^2} J_{2n}(xy) I_{2n}(2\alpha xy) =$$

$$= -\frac{1}{4\alpha^3}\sum_{n=1}^{\infty} n\left[(1+2n) I_{2n}^{1/2}(q,c) + \frac{(4\alpha^2-1)}{8\alpha} I_{2n}^{3/2}(q,c) - \frac{1}{2} I_{2n+1}^{3/2}(q,c)\right];$$

$$I_{12} = 2\sum_{n=1}^{\infty}\int_{0}^{\infty} dy\, e^{-y^2\left(\frac{1}{2}+\alpha\right)}\int_{0}^{\infty} dx\, x^5 e^{-2\alpha x^2} J_{2n}(xy) I_{2n}(2\alpha xy) =$$

$$= \frac{1}{8\alpha^3}\sum_{n=1}^{\infty}\left[(1+n)(1+2n) I_{2n}^{1/2}(q,c) + (1+n)\frac{(4\alpha^2-1)}{4\alpha} I_{2n}^{3/2}(q,c) + \right.$$

$$\left. + \frac{(4\alpha^2-1)^2}{128\alpha^2} I_{2n}^{5/2}(q,c) - \frac{(4n+5)}{4} I_{2n+1}^{3/2}(q,c) - \frac{(4\alpha^2-1)}{16\alpha} I_{2n+1}^{5/2}(q,c) + \frac{\alpha}{4} I_{2n+2}^{5/2}(q,c)\right];$$

$$I_{13} = -4\sum_{n=1}^{\infty}\int_{0}^{\infty} dy\, y\, e^{-y^2\left(\frac{1}{2}+\alpha\right)}\int_{0}^{\infty} dx\, x^4 e^{-2\alpha x^2} J_{2n}(xy) I_{2n+1}(2\alpha xy) =$$

$$= -\frac{1}{4\alpha^2}\sum_{n=1}^{\infty}\left[(n+1) I_{2n}^{3/2}(q,c) + \frac{(4\alpha^2-1)}{16\alpha} I_{2n}^{5/2}(q,c) - \frac{(2n+3)}{4} I_{2n+1}^{3/2}(q,c) - \left(\frac{1}{4} + \frac{4\alpha^2-1}{32\alpha^2}\right) I_{2n+1}^{5/2}(q,c) + \right.$$

$$\left. + \frac{1}{8\alpha} I_{2n+2}^{5/2}(q,c)\right]; q = \frac{1+4\alpha+4\alpha^2}{8\alpha}, c = \frac{1}{2}, k = \frac{1}{\sqrt{2}}\left(1 - \frac{q}{\sqrt{q^2+c^2}}\right)^{1/2}. \quad (A.6)$$

# References

[1] I.V. Lerner and Yu.E. Lozovik, Zh.Eksp. Teor.Fiz. 80, 1488(1981) [Sov.Phys. – JETP 53, 763, (1981)].

[2] A.B. Dzyubenko and Yu.E. Lozovik, Fiz. Tverd. Tela (Leningrad) 25, 1519 (1983); 26, 1540 (1984) [Sov. Phys. Solid State 25, 874 (1983); 26, 938 (1984)]; J. Phys. A 24, 415 (1991).

[3] D. Paquet, T.M. Rice, and K. Ueda, Phys. Rev. B 32, 5208 (1985), https//doi.org/10.1103/PhysRevB.32.5208.

[4] S.A. Moskalenko, M.A. Liberman, D.W. Snoke, and V.V. Botan, Phys. Rev. B 66, 245316 (2002), https://doi.org/10.1103/PhysRevB.66.245316.

[5] C. Kallin and B.I. Halperin, Phys. Rev. B 30, 5655 (1984), https:/doi.org/10.1103/PhysRevB.30.5655.

[6] J.P. Eisenstein, A.H. Macdonald, Nature 432, 691–694 (2004), doi: 10.1038/nature03081.

[7] A. Wojs, A. Gladysiewicz, D. Wodzinski, and J.J. Quinn, Can. J Phys. 83(10), 1019–1028 (2005), https://doi.org/ 10.1139/p05-046.

[8] A. Wojs, Phys. Rev. B 63, 125312 (2001), https://doi.org/10.1103/PhysRevB.63.125312.

[9] A. Wojs and P. Hawrylak, Phys. Rev. B 56, 13227 (1997), https://doi.org/10.1103/PhysRevB.56.13227.

[10] A. Wojs, J.J. Quinn, Solid State Comm. 100, 1, 45–49 (1999), https://doi.org/10.1016/S0038-1098(99)00004-6.

[11] M.A. Olivares-Robles and S.E. Ulloa, Phys. Rev. B 64, 115302 (2001), https://doi.org/10.1103/PhysRevB.64.115302.

[12] Yu.A. Bychkov and E.I. Rashba, Zh. Eksp. Teor. Fiz. 85, 1826–1846 (1983).

[13] E.V. Dumanov, I.V. Podlesny, S.A. Moskalenko, V.A. Liberman, Physica E, 88, 77–86, (2017).

[14] F.D.M. Haldane, Phys. Rev. Lett. 51, 605 (1983), https://doi.org/10.1103/PhysRevLett.51.605.

[15] M.A. Liberman and A.V. Petrov, Physica Scripta 57, 573 (1998).

[16] Kalman Varga, Few–Body Syst. 47, 65–71 (2010).

[17] T. Baar, M. Bayer, A.A. Gorbunov and A. Forchel, Phys. Rev. B 63, 153312 (2001).

[18] S.A. Moskalenko, P.I. Khadzhi, I.V. Podlesny, E.V. Dumanov, M.A. Liberman, I.A. Zubac, Metastable Bound States of the Two-Dimensional Bi-magnetoexcitons in the Lowest Landau Levels Approximation, Semiconductors, (2018), accepted.

[19] S.A. Moskalenko, P.I. Khadzhi, I.V. Podlesny, E.V. Dumanov, M.A. Liberman, I.A. Zubac, Matrix elements of the optical transitions from the biexciton to para and ortho magnetic excitons, unpublished.

[20] D.I. Blokhintzev, The principles of the quantum mechanics. Moscow, p. 390, 1976, (in Russian).

[21] Yu.P. Kravchenko and M.A. Liberman, Physical Review A 56 (1997) R2510-R2513. https://doi.org/10.1103/PhysRevA.56.R2510.

[22] Yu.P. Kravchenko and M.A. Liberman, Phys. Rev. A 57 (1998) 3403-3418. https://doi.org/10.1103/PhysRevA.57.3403.

[23] Yu.P. Kravchenko and M.A. Liberman, JETP Letters 67 (1998) 429-433. https://doi.org/10.1134/1.567686.

[24] V.B. Timofeev and A.V. Chernenko, Pis'ma Zh. Eksp. Teor Fiz. 61 (1995) 603 [JETP Lett. 61 (1995) 617].

[25] Granino A. Korn, Theresa M. Korn, Mathematical handbook for scientist and engineers, McGraw-Hill Book Company, N. Y., 1968.

[26] I.S. Gradshteyn and I.M. Ryzhik. Table of Integrals, Series, and Products, Moscow, Nauka, 1971, 1108 pages.

[27] A.P. Prudnikov, Yu.A. Brychkov, O.I. Marichev, Integrals and Series: Special functions, CRC Press, 1986, 750 pages.